\documentclass[aps,pra,twocolumn,floats,showpacs]{revtex4}
\usepackage{epsfig}
\usepackage{amsmath}
\usepackage[dvips]{color}
\usepackage{amssymb}
\usepackage{amsopn}
\usepackage{amstext}
\usepackage{amsbsy}
\usepackage{amscd}
\usepackage{amsxtra}
\usepackage{setspace}
\usepackage{bm}
\usepackage{subfigure}

\newcommand{\ra}{\rangle}

\begin{document}
\title{Constructing  Entanglers in 2-Players--N-Strategies  Quantum Game} 

\author{Y. Avishai} 
\affiliation{Department of Physics and the Ilse Katz Center
for Nano-Science, Ben-Gurion University, Beer-Sheva 84105, Israel} 

\begin{abstract}
In quantum games  
based on 2-player--$N$-strategies classical games, each player has a quNit (a normalized vector in 
an $N$-dimensional Hilbert space ${\cal H}_N$) upon which he applies his strategy (a matrix $U \in$ SU(N)). The players draw their payoffs 
from a state $|\Psi \ra=J^\dagger U_1 \otimes U_2 J|\Psi_0 \ra \in {\cal H}_N \otimes {\cal H}_N $.  Here $|\Psi_0 \ra$ 
and $J$  (both determined by the game's referee) are respectively an {\it unentangled} 2-quNit (pure) 
state and a unitary operator
such that $|\Psi_1 \ra \equiv J|\Psi_0 \ra \in {\cal H}_N \otimes {\cal H}_N$ is {\it partially entangled}. 
The existence of pure strategy Nash equilibrium in the quantum game is 
intimately related to the degree of entanglement of $|\Psi_1 \ra$. 
Hence, it is  practical to design the entangler $J=J(\beta)$ to be dependent on a {\it single}
real parameter $\beta$ that controls the degree of entanglement of $|\Psi_1 \ra$, 
such that its von-Neumann entropy $ {\cal S}_N(\beta)$ is continuous 
and obtains {\it any value} in $[0, \log N]$. Moreover, an efficient control of $ {\cal S}_N(\beta)$ is 
possible only if $|\Psi_1 \ra$ appears in a Schmidt decomposed form.
Designing $J(\beta)$ for $N=2$
is quite standard. 
Extension to $N>2$ is not obvious, and here we
 suggest an algorithm to achieve it. 
\end{abstract}


\maketitle
The theory of quantum games is an evolving discipline that, similar to quantum information, 
explores the implications of quantum mechanics to fields outside physics proper, such as 
economics, finance, auctions, gambling {\it etc.}\cite{Lev,Meyer,EWL,Flitney,Piotrowski1,Landsburg1,Iqbal}. 
One way of constructing a quantum game is to start from a 
standard (classical) game and to ``quantize" it by formulating appropriate rules and letting
the players employ quantum tools such as qubits and quantum strategies (gates). 
This procedure has  been applied on
classical strategic games that describe an interactive decision-making in which each
player chooses his strategy only once, and all
choices are taken simultaneously. A simple example is a
quantum game based on 2-player--2-strategies classical game usually defined by a game table 
(for example, the prisoner dilemma). We refer to it briefly as a 2-2 game. 
There is an extensive work on the quantized version of classical strategic $2-2$ games, 
most of them are based on the protocol specified in Ref.~\cite{EWL}. It requires application of
an  entanglement (unitary) operator $J(\beta)$ (where $\beta$ is a real parameter),
that acts on a non-entangled 2-qubit (pure) state resulting in an entangled state whose degree of 
entanglement is measured by its von-Neumann entropy ${\cal S}_2(\beta)$. 
A desired property of $J(\beta)$ is that 
${\cal S}_2(\beta)$ is a continuous function of $\beta$ that
varies (preferably monotonically) between $0$ (no entanglement) and $\log 2$ (maximal entanglement). 
The reason for exploring partially entangled 2-qubit states  is that the existence of pure strategy Nash equilibrium in the 2-2 quantum game crucially 
depends on the degree of entanglement (see below). Controlling the entropy by a {\it single parameter} such that all 
values between 0 and $\log 2$ are obtained is referred to here as {\it single parameter completeness}. 
Another practical property required from $J(\beta)$ is that it can easily be constructed 
from the {\it classical strategies}. In a 2-2 game, the classical strategy of a player is $i \sigma_y  \in $ SU(2), and 
an appropriate construction is then $J(\beta)=e^{-i \tfrac{\beta}{2} \sigma_y \otimes \sigma_y}$.  Its action on 
an unentangled 2-qubit state (e.g $|00 \ra$) yields,
\begin{equation} \label{1}
J(\beta)|00 \ra=e^{-i \tfrac{\beta}{2} \sigma_y \otimes \sigma_y}|00 \ra=\cos \tfrac{\beta}{2}|00\ra-i \sin \tfrac{\beta}{2}|11 \ra,
\end{equation} 
where the left (right) factor in the Kronecker product refers to player 1 (2). In this way, $|\Psi_1 \ra=J(\beta)|00 \ra$ 
appears in a {\it Schmidt decomposed form}, enabling an easy computation of 
the corresponding entanglement entropy of the 2-qubit state on the RHS as, 
\begin{equation} \label{Sbeta}
{\cal S}_2(\beta)=-\left (\cos^2 \tfrac{\beta}{2} \log \cos^2 \tfrac{\beta}{2} +  \sin^2 \tfrac{\beta}{2} \log  \sin^2 \tfrac{\beta}{2} \right).
\end{equation} 
Thus, ${\cal S}_2 (\beta)$ is a continuous function of $\beta$ and 
gets all values in $[0, \log 2]$ namely, $J(\beta)$ as defined in Eq.~(\ref{1}) satisfies 
single parameter completeness. Other properties (of less significance) are that ${\cal S}_2 (\beta)$
is periodic  with period $\pi$, symmetric about the 
mid-point $\tfrac{ \pi}{2}$, with ${\cal S}_2(0)=0$ and ${\cal S}_2(\tfrac{ \pi}{2})=\log 2$. 
  
The entangler defined in Eq.~(\ref{1}) has a property referred to as  {\it classical commensurability}, 
  $[J(\beta), \sigma_y \otimes \sigma_y]=0$. Following the rules of the game specified 
in Ref.~\cite{EWL} it means that players in a 2-2 quantum game can, if they wish, 
use their classical strategies as a special case and if they do so, they collect the corresponding 
classical payoffs.  In most cases, however,  the classical strategies do not constitute 
a pure-strategy Nash Equilibrium (NE) (defined below).
Generically, as we explain below, classical commensurability does not hold for $N>2$ (see however Ref.~\cite{Sharif}). 

\begin{figure}
\vspace{0.2in}
\centering\subfigure[]{\includegraphics[width=0.195\textwidth, angle=0]
{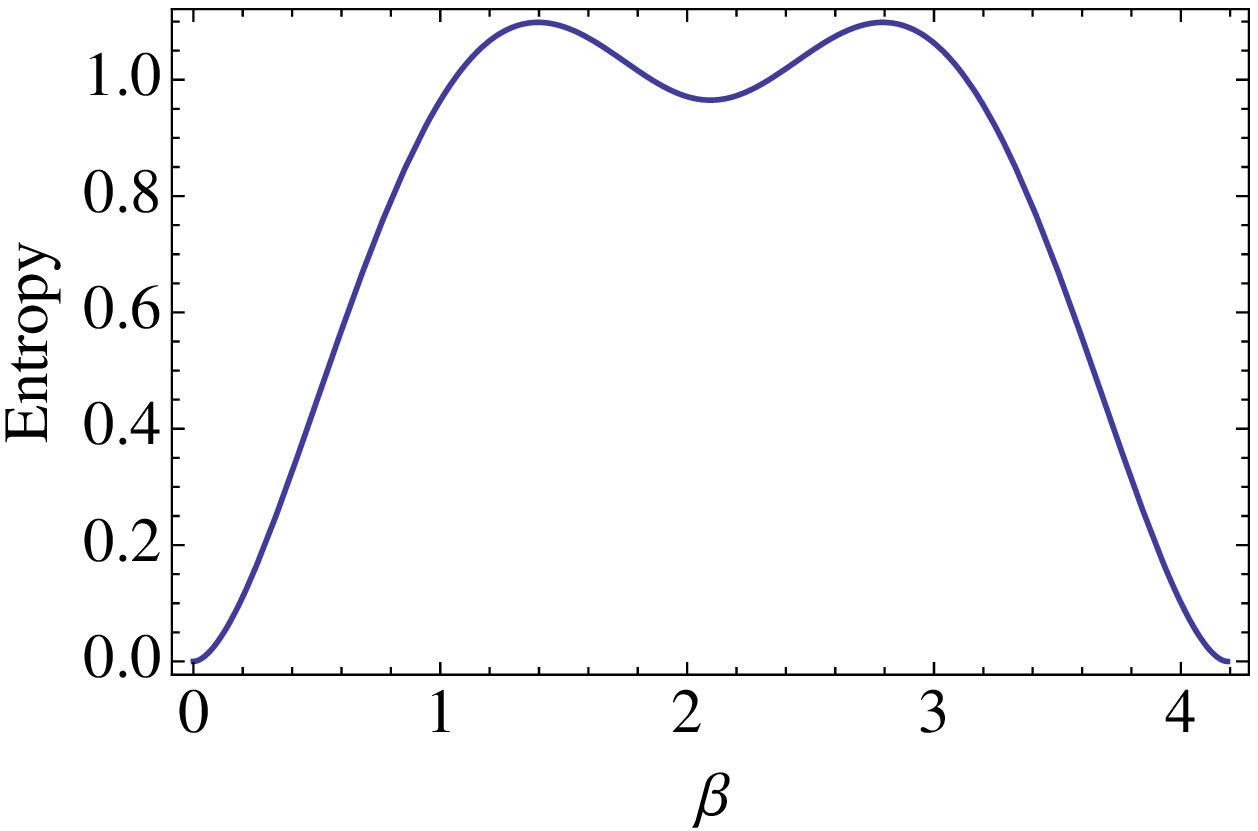}}
\centering\subfigure[]{\includegraphics[width=0.195\textwidth, angle=0]
{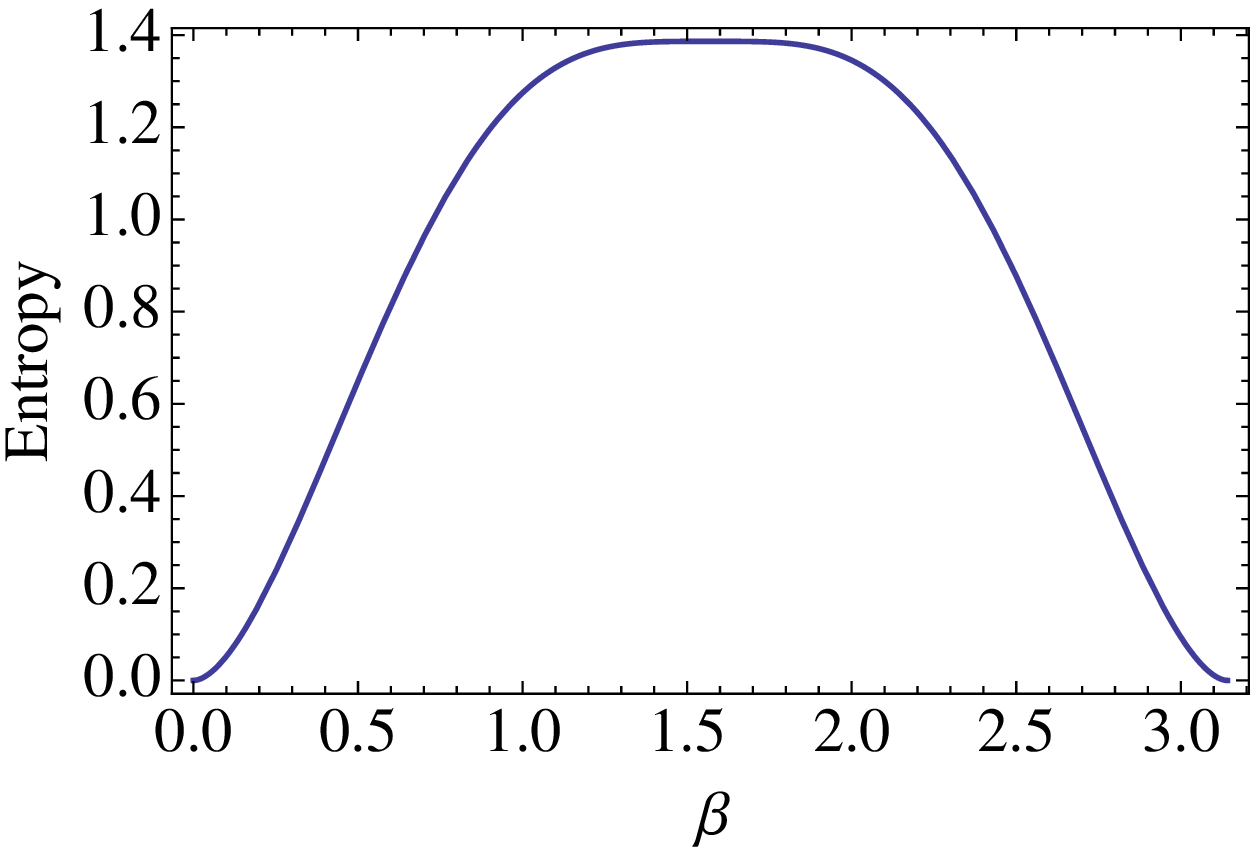}}
\caption{\footnotesize{von Neumann entropy $S_N(\beta)$ defined in Eq.~(\ref{entropyN}), 
for the 2-quNit states  
$J(\beta)|1 1 \ra$ defined in Eq.~(\ref{JbetaN}): (a) N=3, 
(b) N=4.  $S_N(\beta)$ varies continuously reaching all values in the interval $[0,\log N ]$, so that 
single parameter completeness is 
satisfied. Here $S_N(\beta)$ is periodic with period $\tfrac{4 \pi}{N}$. }}
\label{Fig1}
\vspace{-0.28in}
\end{figure}

 In the present work we examine the issue of constructing $J(\beta)$ 
 for  a $2-N$ quantum game based on a 2-players--N-strategies classical 
 game. 
 We  suggest a natural extension of Eq.~(\ref{1}) for constructing an operator $J(\beta)$ that 
 turns a non-entangled 2-quNit states into an entangled one. For $N=3,4$  
 the corresponding von-Neumann entropy ${\cal S}_N(\beta)$ varies continuously between
0 and $\log N$ so that single parameter completeness is satisfied. 
Unfortunately, this method does not work for $N>4$ because in that case ${\cal S}_N(\beta)<\log N$. 
To alleviate this deficiency we suggest another method 
(albeit less intuitive) to design $J(\beta)$ that satisfies single parameter completeness for any $N$.
  In what follows we will 
 first introduce the classical 2-3 game using quantum information language and  formulate its quantum 
 version (extension to $N>3$ is straightforward). Most of this introductory exposition is well established and is 
 included here merely for self consistence.  
Then, in the second step, we shall address the issue of constructing $J(\beta)$.    \\
 \ \\
 \underline{Two-Players--Three-Strategies Classical Games: Trits}\\
Consider a two-players classical game with three strategies for each player. 
For example, two prisoners may have three options, marked as three values of a trit 
C=$|1 \ra$,S=$|2 \ra$ and D=$|3 \ra$ for Confess, Stay quiet  or Don't confess.
The two prisoner ``system" can be found in  
nine two-trit states  $|i j \ra, \ \ i=1,2,3 \ \ j=1,2,3$,
corresponding to the nine entries of the game table.
The protocol of the classical game with 2-players and 3-strategies is as follows: 
The referee (judge) calls the players (prisoners) 
and tells them he assumes that they are in an initial 
two-trit state $|1 1 \ra$ meaning (C,C) namely both confess. He then asks them to 
decide whether to leave their respective trit state as it is on $|1 \ra$ or to change it either to $|2 \ra$ (meaning S) or to $|3 \ra$ (meaning D). These replacement operations (specified explicitly below) are the players classical strategies. 
If the initial state suggested by the referee is $|11 \ra$ 
the strategies of the two players include ${\bf 1}_3$ ( 
leaving the trit at $|1 \ra$ as it is), 
$S_{12}$ (swapping of $|1 \ra $ and $|2 \ra$ namely, replacing C by S) and $S_{13}$ (swapping $|1 \ra$ and $|3 \ra$ namely replacing C by D). 
These three operations generate the group ${\bf S}_3$ of permutations of three elements $\binom {123}{ijk}$. Explicitly,\\ 
${\bf 1}_3$=(123), \ $S_{12}$=(213), $S_{13}$=(321), $S_{13}S_{12}S_{13}$=$S_{23}$=(132)\\
 $S_{13}S_{12}$=$S_{312}$=(312), \ $S_{12}S_{13}$=$S_{231}$=(231). 
 \\
 We shall indicate below that a quantum strategy is a gate represented by an SU(3) matrix. A reasonable 
 requirement for the procedure of ``quantizing" a classical game is that
 the classical strategies obtain as special case of the quantum ones. For that purpose  
we need to construct a representation of the permutation group ${\bf S}_3$ 
in terms of unitary matrices with unit determinant.
This can be achieved by choosing, 
 \begin{eqnarray}
&& {\scriptsize%
    S_{12} \mbox{=} \begin{bmatrix} 0 & -1 & 0 \\ -1 & 0 & 0 \\ 0 & 0 & -1 \end{bmatrix} \  S_{13} \mbox{=} \begin{bmatrix} 0 & 0 & -1 \\ 0 & -1 & 0 \\ -1 & 0 & 0 \end{bmatrix} \  S_{23} \mbox{=} \begin{bmatrix} -1 & 0 & 0 \\ 0 & 0 & -1 \\ 0 & -1 & 0 \end{bmatrix} }
 \nonumber \\
&& {\scriptsize%
    S_{312} = \begin{bmatrix} 0 & 0 & 1 \\ 1 & 0 & 0 \\ 0 & 1 & 0 \end{bmatrix} \ \ S_{231} = \begin{bmatrix} 0 & 1 & 0 \\ 0 & 0 & 1 \\ 1 & 0 & 0 \end{bmatrix} \ \ {\bf 1}_3 = \begin{bmatrix} 1 & 0 & 0 \\ 0 & 1 & 0 \\ 0 & 0 & 1 \end{bmatrix} }
\label{SUC}
\end{eqnarray}
For example, suppose player 1 and 2 choose respective strategies $s_1=S_{12}$ and $s_2=S_{13}$. 
This brings the system into a state $s_1 \otimes s_2|11 \ra=|23\ra=|\mbox{SD}\ra$. Then the respective payoff of 
player $k, \ k=1,2$ will be $u_k(s_1,s_2)= u_k(\mbox{S,D})=u_k(2,3)$ where $u_k(i,j)$ is the payoff of player $k$ 
at entry $(i,j)$ of the game table.  \\
\underline{The analogous Quanum Game:}
We now briefly explain the 
structure of the corresponding quantum game. 
Its main ingredients are qutrits, quantum strategies, and entanglement operations. 
Both versions use the same game table but the 
payoff rules are somewhat different. \\
 \noindent
 \underline{1 and 2 qutrit states} \\
Consider the three dimensional Hilbert space ${\cal H}_3$ with orthonormal basis vectors $|1 \ra, |2 \ra$ and $|3 \ra$. 
 A qutrit is a vector $|\psi \ra=v_1|1\ra+v_2|2 \ra + v_3|3 \ra \in {\cal H}_3$ of unit norm, $|v_1|^2+|v_2|^2+|v_3|^2=1$. 
 \noindent
 A general 2-qutrit state is a normalized vector 
 in ${\cal H}_3 \otimes {\cal H}_3$, 
 \begin{equation} \label{2qutritgen}
 |\Psi \ra=\sum_{i,j=1}^3 v_{ij}|i j \ra , \ \  v_{ij} \in \mathbb{C}, \ \ \sum_{i,j=1}^3 |v_{ij}|^2=1~.
 \end{equation}
A {\it maximally entangled two-qutrit state} is written as, 
\begin{equation} \label{Max_2qutrit}
|\Psi \ra_{\mathrm {ME}}=\frac{1}{\sqrt{3}} (u_1|1 1 \ra+u_2|22 \ra+u_3|3 3 \ra), \ \ u_i \in \mathbb{C}, \ \ |u_i|=1~,
\end{equation} 
given in a Schmidt decomposed form. Its entanglement degree is measured  by the von Neumann entropy \\
  $\hphantom{1234567} {\cal S}_3=-\tfrac{1}{3} \sum_{i=1}^3 |u_i|^2 \log (|u_i|^2/3)=\log 3.$\\
 \noindent
 \underline{ Quantum Strategies}\\
 A strategy of a player in a 2-3 quantum game is an SU(3) matrix by which he operates on his qutrit (that is a quantum gate). 
A strategy $ U({\bm \gamma})\in$SU(3) depends on eight Euler angles ${\bm \gamma} \equiv \{ \alpha_1,\alpha_2,\ldots,\alpha_8 \}$. 
The explicit expression of $ U({\bm \gamma})$ in terms of Gellman matrices $\{ \lambda_{m} \}, \ \ (m=1,2,\ldots,8)$ is well known. 
For quantum game theory, a practical parametrization of $U({\bm \gamma})$ is suggested in Ref.\cite{Sharif}. \\
 \noindent
\underline{2-3 Quantum Game Procedure}\\
The referee suggests an initial non-entangled 
 two-qutrit state $|\Psi_0 \ra$ ({\it e.g} the analog of the 
 classical two trit state $|\Psi_0 \ra\mbox{=}|11 \ra$).   
 Before letting  the players apply their quantum strategies, the referee operates on $|\Psi_0 \ra$ with a unitary operator 
 $J(\beta)$ such that $|\Psi_1 \ra=J(\beta)|\Psi_0 \ra$ is entangled (otherwise the game remains classical). 
 Construction of the operator $J(\beta)$  (our central goal) is detailed below. At this stage of the game, 
 the players apply their respective strategies 
 $U_1 \otimes U_2$. Finally, the referee applies the operator $J^\dagger (\beta)$ leading to the final 
 state 
 \begin{equation} \label{24}
 |\Psi \ra\mbox{=}\overbrace{J^\dagger(\beta)}^{\mbox {referee}} \overbrace{U({\bm \gamma}_1) \otimes U({\bm \gamma}_2)}^{\mbox { players}} 
 \overbrace{J(\beta)|11 \ra}^{\mbox {referee}}\mbox{=} \displaystyle \sum_{i,j=1,3} v_{ij}|ij \ra,
  \end{equation}
  where ${\bm \gamma}_k$ is the octet of 8 Euler angles defining the SU(3) matrix $U({\bm \gamma}_k)$ 
  (that is the strategy of player $k=1,2$). 
  The payoff  $P_k$ of player $k=1,2$  is given by, 
  \begin{equation} \label{PQ}
 P_k({\bm \gamma}_1,{\bm \gamma}_2)= \displaystyle \sum_{i,j=1,3} u_k(i,j)|v_{ij}({\bm \gamma}_1,{\bm \gamma}_2)|^2, \ \ k=1,2~,
 \end{equation} 
 where $u_k(i,j)$ are the payoffs at entry $(i,j)$ of the {\it classical} game table.  
 Like in the classical game, each player choses a strategy with 
 the goal of maximizing his payoff. 
 \newpage
 \noindent
 \underline{Pure Strategy Nash Equilibrium (NE)}\\
 Because the set of 8 Euler angles ${\bm \gamma}$ uniquely determines the player's strategy $U({\bm \gamma}) \in$ SU(3),
 a pure strategy NE in the 2-3 quantum game is a pair of strategies $({\bm \gamma}_1^*,{\bm \gamma}_2^*)$ (each represents 8 angles), such that 
\begin{eqnarray} \label{34}
&& P_1({\bm \gamma}_1,{\bm \gamma}_2^*) \le P_1({\bm \gamma}_1^*,{\bm \gamma}_2^*) \ \forall \ \ {\bm \gamma}_1,
 \nonumber \\
&& P_2({\bm \gamma}_1^*,{\bm \gamma}_2) \le P_2({\bm \gamma}_1^*,{\bm \gamma}_2^*) \ \forall \ \ {\bm \gamma}_2 .
 \end{eqnarray}
 The question of whether pure strategy NE exists in 2-2 quantum game, and its relation to the degree 
 of entanglement (controlled by $\beta$) has been discussed in numerous works\cite{Landsburg2,Hayden,Du,Du1,Flitney1,Flitney2, Avishai}. 
In brief,  if there is NE in the classical game that is {\it not Pareto efficient}\cite{Osborn}, then 
there is a critical value $\beta_c$ above which there is no pure strategy NE in the quantum game.  
As $\beta$ approaches $\beta_c$ from below, the respective payoffs in the quantum 
game at NE approach the Pareto point of cooperation\cite{Du,Avishai}.
This is the main reason why, right from the onset, 
we stressed the relevance of partially  entangled 2-quNit states where ${\cal S}_N(\beta) < \log N$.\\
\noindent
\underline{Absence of classical commensurability}\\
We now explain why, in a 2-3 quantum game there is no classical commensurability\cite{Sharif}. 
Recall that classical commensurability means that if the players use classical strategies they 
respectively get their classical payoffs. For a classical strategy ${\bf \bar{\gamma}}$ we have $U({\bf \bar{\gamma}}) \in {\bf S}_3$.  
From Eq.~(\ref{24})  it means that $J$ should commute with all outer products of the classical strategies.  
If the initial  state is $|\Psi_0\ra=|11 \ra$ the 9 outer products of classical strategies 
are $s_1 \otimes s_2$, where $s_k \in \{ {\bf 1}_3,S_{12},S_{13} \}$, (see Eq.~(\ref{SUC})). 
Classical commensurability then requires,
\begin{equation} \label{ccqutrits}
 [J, S_{12} \otimes S_{13}]=[J, S_{13} \otimes S_{12}]=0.
\end{equation}
This is possible only if $J$ is a function of $A \otimes A$ where $A$ is a $3 \times 3$ matrix
 satisfying $[A,S_{12}]=[A,S_{13}]=0$, and $A$ is not just a multiple of ${\bf 1}$. 
 But this is impossible because $S_{12}$ and $S_{13}$ generate an irreducible representation of the permutation group 
 $S_3$ and hence, according to Schure's lemma $A$ is just a multiple of ${\bf 1}_3$. These arguments naturally 
 hold for any $N>2$. \\
 \noindent
\underline{Designing $J(\beta)$}\\
The main result of the present study concerns the analysis and construction of an entanglement 
 operator $J(\beta)$ for $N>2$. We carry it out for $N=3$ and then extend it straightforwardly to any $N$. 
 For $N=3$ we require that 
$J(\beta) |11 \ra$  yields an entangled 2-qutrit state with $\beta$ specifying the degree of entanglement 
that achieves {\it any value} between 0 and $\log 3$. 
Following the 2-2 game framework specified in Eq.~(\ref{1}),
we try to construct $J$ by exponentiating a combination of {\it classical strategies}. 
In order to get the 
``diagonal" 2-qutrit states $|22 \ra$ and $|33 \ra$ from the qutrit state $|11 \ra$ we have to operate on $|11 \ra$ 
with $Z \equiv  S_{12} \otimes S_{12}+S_{13} \otimes S_{13}$. Therefore, we define 
\begin{equation} \label{Jbeta}
J(\beta)=e^{i \tfrac{\beta}{2}  Z}=e^{i \tfrac{\beta}{2} (S_{12} \otimes S_{12}+S_{13} \otimes S_{13})}~.
\end{equation}
Calculation of the exponent yields 
\begin{equation} \label{qutrits_J1}
J(\beta)|11 \ra \mbox{=}\frac{1}{3} \left [ (2 e^{-i \tfrac{\beta}{2}}\mbox{+}e^{i \beta}) |11 \ra \mbox{+}(e^{ i \beta} \mbox{-}e^{-i\tfrac{\beta}{2}})(|22 \ra+|33 \ra) \right ]~. 
\end{equation} 
Maximal entanglement obtains when the absolute values of all three coefficients are equal, namely,
\begin{equation} \label{JME}
|2 e^{-i \tfrac{\beta}{2}}\mbox{+}e^{i \beta}|=|e^{ i \beta} \mbox{-}e^{-i\tfrac{\beta}{2}}| \ \ \Rightarrow \ \ \beta=\frac{4 \pi}{9},\frac{8 \pi}{9}~.
\end{equation}
Here ${\cal S}_3(\beta)$ raises monotonically from $0$ to its first maximum 
$\log 3$, hence  we have found the desired entanglement operator that satisfies single parameter completeness. 
Fig.~\ref{Fig1}(a) displays the 
von Neumann entropy ${\cal S}_3(\beta)$ of the entangled 2qutrit state (\ref{qutrits_J1}) as function of $\beta$. 
Here again it possesses other properties, namely, 
${\cal S}_3(\beta)$ is a periodic function of $\beta$ with period $\tfrac{4 \pi}{3}$ and it is symmetric about the 
mid-point $\tfrac{2 \pi}{3}$ where it has a local {\it minimum}. It has two maxima for $\beta=\frac{4 \pi}{9},\frac{8 \pi}{9}$ where  
it equals $\log 3$. 
Inspecting Eq.~(\ref{Jbeta}), we see that, in quantum games, the entanglement is not obtained in terms 
of spin rotations but, rather, in terms of permutation exponentials 
that are SU(3) matrices (for $N=2$ these are the same). \\
\underline{Extension to arbitrary N:} Let $S_{ij} \in {\bf S}_N$ denote the $N \times N$ matrix  representing the 
permutation $i \leftrightarrow j$ and let $|i j \ra, \ i,j=1,2,\ldots ,N$ be an unentangled 2-quNit state. To get an entangled state 
from $|1 1 \ra$ we define and assert that
\begin{eqnarray} \label{JbetaN}
&& J(\beta)|1 1 \ra=e^{i \tfrac{\beta}{2} \sum_{j=2}^N S_{1j} \otimes S_{1j}} |11\ra = \frac{e^{-i \tfrac{\beta}{2}}}{N} \nonumber \\
&&  \left [  \left ( e^{i \tfrac{N \beta}{2}} \mbox{+}N \mbox{-}1 \right )|11 \ra 
\mbox{+}\left ( e^{i \tfrac{N \beta}{2}} \mbox{-} 1 \right ) \sum_{j=2}^N  |j j \ra \right ]. 
\end{eqnarray} 
To proceed, let us define the absolute value squared of the coefficients,
\begin{equation} \label{f1f2}
f_1(\beta)=\frac{| e^{i \tfrac{N \beta}{2}} \mbox{+}N \mbox{-}1|^2}{N^2}, \ f_2(\beta)=\frac{| e^{i \tfrac{N \beta}{2}} \mbox{-}1|^2}{N^2}~.
\end{equation}
It is easy to verify that 1) $0 \le f_i(\beta) \le 1$ 2)  $f_1(\beta)+(N-1) f_2(\beta)=1 \ \forall  \ \beta$ and 3) $f_i(\beta)$ is periodic 
with period $\tfrac{4 \pi}{N}$ and symmetric about the mid-point $\tfrac{2 \pi}{N}$. The entanglement entropy is
\begin{equation} \label{entropyN}
{\cal S}_N(\beta)=-[f_1(\beta) \log f_1(\beta)+(N-1) f_2(\beta) \log f_2(\beta)~.
\end{equation} 

Maximal entanglement ${\cal S}_N(\beta)=\log N$ obtains for $\beta_0$ that is the solution of the equality 
\begin{equation}\label{beta0}
\frac{| e^{i \tfrac{N \beta_0}{2}} \mbox{+}N \mbox{-}1|^2}{N^2}=\frac{| e^{i \tfrac{N \beta_0}{2}} \mbox{-}1|^2}{N^2}=\tfrac{1}{N}. 
\end{equation}
For $N=3$ the two solutions are specified in Eq.~(\ref{JME}). For $N=4$, there is a single solution at $\beta_0=\tfrac{\pi}{2}$, as shown in Fig. \ref{Fig1}(b). 
Thus, for $N=3,4$ we have achieved our goal of constructing an entanglement operator $J(\beta)$  
such that the degree of entanglement 
$S_N(\beta)$ of the 2-quNit state $J(\beta)|1 1 \ra$  
varies continuously reaching all values in the interval $[0,\log N ]$, so that 
single parameter completeness is satisfied. 

\begin{figure}
\vspace{0.2in}
\centering\subfigure[]{\includegraphics[width=0.195\textwidth, angle=0]
{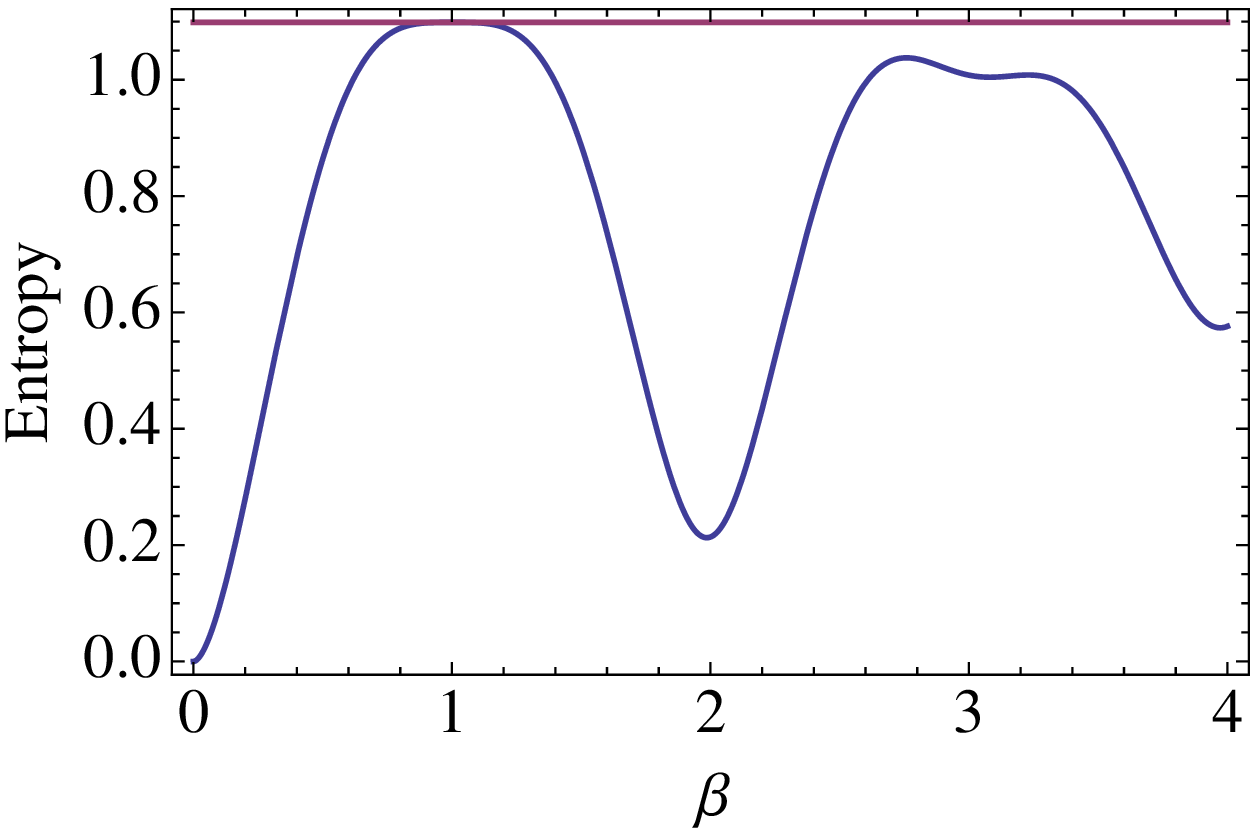}}
\centering\subfigure[]{\includegraphics[width=0.195\textwidth, angle=0]
{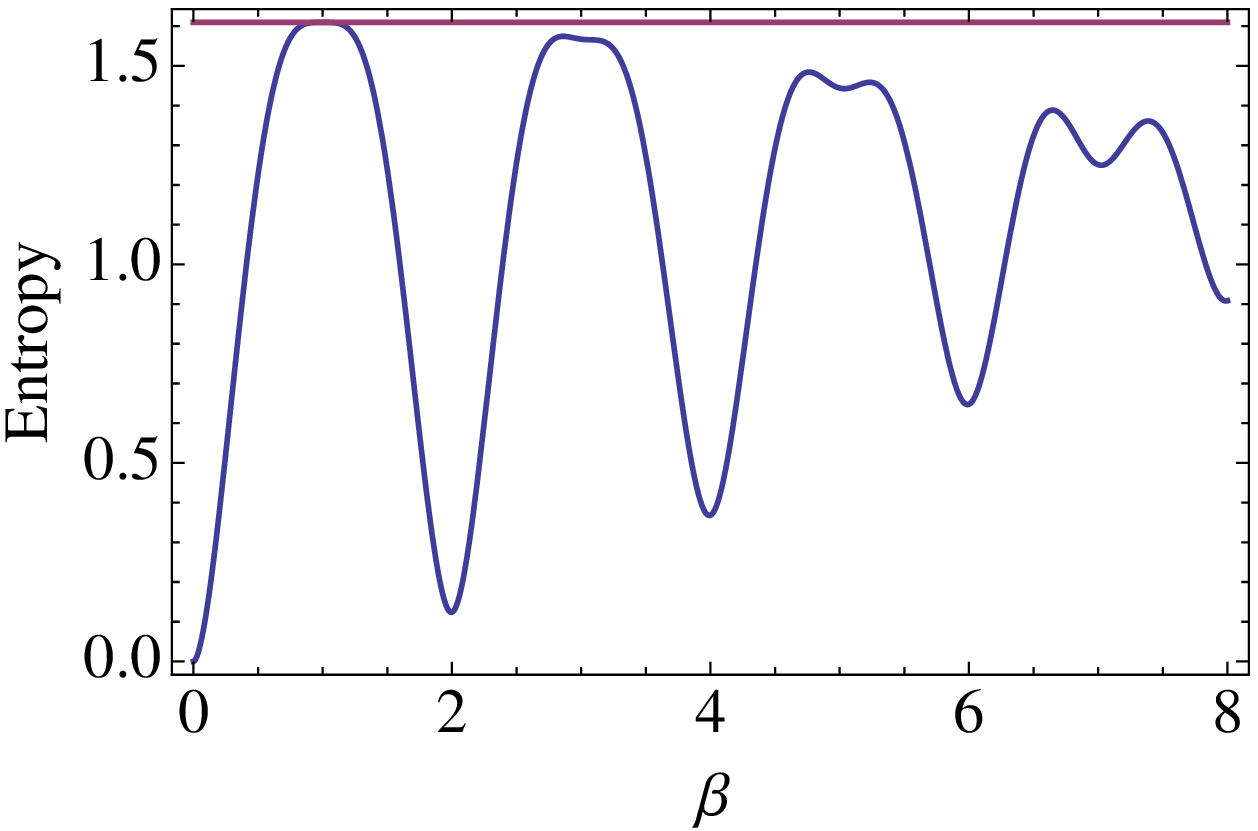}}
\caption{\footnotesize{von Neumann entropy for 2-quNit state $|\Psi_1\ra= J(\beta)|11 \ra$, 
where $J(\beta)$ is defined in Eq.~(\ref{JbetaNU}). (a),(b) correspond to $N=3,5$.   
By construction, ${\cal S}_N(0)=0$ and ${\cal S}_N(1)=\log N$ (horizontal lines). Since ${\cal S}_N(\beta)$ 
is a continuous function of $\beta$ it reaches any value in the interval $[0, \log N]$. That is, 
single parameter completeness is achieved for $N>4$. 
} }
\label{Fig2}
\vspace{-0.2in}
\end{figure}

For $N>4$ there is no solution $\beta_0$ of Eq.~(\ref{beta0}), and maximal entanglement is not achieved. 
It might be argued that 2-N quantum games with $N>4$ are much rarer than those with smaller $N$ but we believe 
that the construction of $J(\beta)$ that satisfies single parameter completeness 
also for $N>4$ is useful in other areas, (outside the ballpark of quantum game theory), so 
we carry it out for the sake of completeness. 
\ \\
The method suggested here is not based on permutation exponentials as in Eq.~(\ref{Jbeta}). It consists of the following steps.
\begin{enumerate}
\item{} Assume a lexicographic order of the $N^2$ basis states $\{ |ij \ra \}$ such that the diagonal states $\{ | ii \ra \}$ appear 
in the first $N$ places.  Choose a unitary $N^2 \times N^2$ matrix of the form $\binom {R \ 0}{0 \ I}$ 
where $R$ is an $N \times N$ 
uniray matrix with equal first column elements $R_{i1}=\tfrac{1} {\sqrt{N}}$ and 
$I$ is the $N(N-1) \times N(N-1)$ unit matrix.
The problem is then reduced to the $N$ dimensional subspace spanned by $\{ |ii\ra \}$. 
By construction, 
$$R|11 \ra=\frac{1}{\sqrt{N}} \sum_{i=1}^N|ii\ra,$$
that is a maximally entangled state.
\item{} Diagonalize  $R$ as $R=V \Lambda V^{-1}$ where $V$ is the matrix of eigenvectors of $R$, and 
$$ \Lambda \mbox{=diag}\{ e^{i \eta_1},e^{i \eta_2},\ldots, e^{i \eta_{N}} \},$$
is the diagonal matrix of (unimodular) eigenvalues of $R$ with eigenphases $\{ \eta_i \}$. 
\item{} Now consider the matrix
\begin{eqnarray} \label{JbetaNU}
&& J(\beta) \equiv V \Lambda(\beta) V^{-1}, \ \mbox{with} \nonumber \\
&& \Lambda(\beta) \equiv \mbox{diag}\{ e^{i \beta \eta_1},e^{i \beta \eta_2},\ldots, e^{i \beta \eta_{N}} \}.
\end{eqnarray}
By construction, $J(0)={\bf 1}_{N \times N}$ and $J(1)=R$. 
Hence, the state $J(\beta)|11\ra=\sum_{i=1}^N [J(\beta)]_{i 1}|ii \ra$ is partially 
entangled. Since it is given in a Schmidt decomposed form, 
the corresponding von Neumann entropy ${\cal S}_N(\beta)$ is easily calculable.  ${\cal S}_N(\beta)$ is
continuous in $[0, \infty)$ with ${\cal S}_N(0)=0$ and ${\cal S}_N(1)=\log N$, namely, 
single parameter completeness is satisfied. Generically, the eigenphases are not rational multiples of $\pi$ so that ${\cal S}_N(\beta)$ is not periodic, 
but this lack of periodicity is of no special significance. 
\end{enumerate} 
\underline{Illustration for $N=5$ :}  
A convenient way to build an appropriate unitary $N \times N$ matrix $R$ is to start from a simple non-singular matrix $A$ 
and then orthogonalize it within the Grahm-Schmidt procedure. For example, 
\begin{equation*}
{\scriptsize%
 A=\begin{bmatrix}1&1&1&1&1\\1&1&0&0&0 \\1&0&1&0&0\\1&0&0&1&0\\1&0&0&0&1 \end{bmatrix}, 
\ \ R=
 \begin{bmatrix}
 \tfrac{1}{\sqrt{5}}&\sqrt{\tfrac{3}{10}}&\sqrt{\tfrac{3}{14}}&\tfrac{3}{2 \sqrt{14}}&\tfrac{1}{2 \sqrt{2}}\\
 \tfrac{1}{\sqrt{5}}&\sqrt{\tfrac{3}{10}}&-\sqrt{\tfrac{3}{14}}&-\tfrac{3}{2 \sqrt{14}}&-\tfrac{1}{2 \sqrt{2}}\\
 \tfrac{1}{\sqrt{5}}&-\sqrt{\tfrac{2}{15}}&2 \sqrt{\tfrac{2}{21}}&-\tfrac{3}{2 \sqrt{14}}&-\tfrac{1}{2 \sqrt{2}}\\
\tfrac{1}{\sqrt{5}}&-\sqrt{\tfrac{2}{15}}&-\sqrt{\tfrac{2}{21}}& \tfrac{5}{2 \sqrt{14}}&-\tfrac{1}{2 \sqrt{2}}\\
\tfrac{1}{\sqrt{5}}&-\sqrt{\tfrac{2}{15}}&-\sqrt{\tfrac{2}{21}}&-\tfrac{1}{ \sqrt{14}}&\tfrac{1}{\sqrt{2}}~.
\end{bmatrix}
}
\end{equation*}
\ \\
\ \\
Proceeding with the list of steps prescribed above we 
can easily construct $J(\beta)$ and compute the von Neumann entropy of the state $|\Psi_1 \ra= J(\beta)|11 \ra$ upon 
which the players apply their strategies according to the game protocol specified in Eq.~(\ref{24}). The result is given in Fig.~\ref{Fig2}.
As explained in the figure's caption, the degree of entanglement is controlled by a single parameter and 
${\cal S}_N(\beta)$ 
is a continuous function of $\beta$  reaching any value in the interval $[0, \log N]$.  Thus we have achieved our goal of 
constructing an entangler $J(\beta)$ that turns a non-entangled 2-quNit state into an entangled one given in a Schmidt decomposed form
with single parameter completeness satisfied. 
\ \\
\ \\
\noindent
\underline{Summary:}
In conclusion, we suggest two methods to design an entanglement operator $J(\beta)$ that turns a non-entangled 
2-quNit state to a partially entangled state whose von Neumann entropy is fully controlled by a single real parameter. 
The first method is intuitively clear and simple, based on exponential of classical strategies, Eq.~(\ref{Jbeta}), 
and results in a the von Neumann entropy, as displayed in Figure \ref{Fig1}. This method 
does not work for $N>4$ because the resulting entropy does not 
reach the maximally entangled value $\log N$. For that reason we suggest another method that is somewhat less 
transparent but works for any $N$. The resulting entropy as function of $\beta$, is displayed in Figure \ref{Fig2}. 
\ \\
{\bf Acknowledgement} \\
I would like to thank Oscar Vollij for his excellent course in (classical) game theory.
Discussions with Eytan Bachmat, Hosho Katsura, Rioichi Shindou, Doron Cohen and Yehuda Band are highly appreciated. 
This work is partially supported by grant 400/2012 of the Israeli Science Foundation (ISF).

\end{document}